\documentclass[prl,aps,twocolumn,showpacs]{revtex4}
\usepackage{graphicx,latexsym}
\usepackage{dcolumn}
\usepackage{amsmath,amssymb,epsf,bm}
\begin{document}
\title{Spin-Hall effects in a Josephson contact}
\author{A.~G. Mal'shukov$^{1,2,3}$, C.~S. Chu$^{2}$}
\affiliation{$^1$Institute of Spectroscopy, Russian Academy of
Sciences, 142190, Troitsk, Moscow oblast, Russia \\
$^2$Department of Electrophysics, National Chiao Tung University,
Hsinchu 30010, Taiwan \\
$^3$National Center for Theoretical Sciences, Physics Division,
Hsinchu 30043, Taiwan}
\begin{abstract}
The Josephson tunneling through a 2D normal contact with the
spin-orbit split conduction band has been studied in the diffusive
regime. Linearized Usadel equations for triplet components of the
pairing function revealed a striking similarity to the equations
of spin diffusion driven by the electric field in normal metals.
Consequently, we predict that the out-of-plane spin-Hall
polarization accumulates towards lateral sample edges and the
in-plane polarization is finite throughout the entire normal
region. At the same time, the spin-Hall current is absent in the
considered case of the stationary Josephson effect.
\end{abstract}

\maketitle

In connection with various spintronic applications, much interest
have been attracted recently to spin-orbit interaction (SOI)
effects on electron transport in normal metals and semiconductors.
This interaction gives rise to fundamental transport phenomena,
such as the spin-Hall effect (SHE) (for a review see \cite{Engel})
and electric spin orientation \cite{Edelstein,Engel}. These
effects represent a direct manifestation  of the spin-orbit
coupling between spin and charge degrees of freedom in electron
transport. At the same time, spin-orbit effects were also
discussed for superconductors. Some works dealt with SFS junctions
\cite{sfs} (F stands for ferromagnet), others considered SNS
\cite{sns}, SN \cite{Edelstein2} systems, or bulk superconductors
\cite{Edelstein2, Gorkov}. As was pointed out in Ref.
\cite{Edelstein2, Gorkov}, SOI leads to admixture of triplet
components to the pairing function. This sort of singlet-triplet
coupling looks similar to the spin-charge coupling in normal
systems. Therefore, one would expect that phenomena closely
related to SHE could manifest themselves in superconductors. At
the first sight on this problem it becomes clear that, at least in
the case of zero voltage across the junction, the spin-Hall
current can not be generated as a linear response to the
superconducting current. The reason is that these currents have
opposite parities with respect to the time inversion, while they
must be equal in the stationary nondissipative superconducting
transport. On the other hand, besides the spin currents, in normal
systems SHE leads to spin accumulation near sample edges.
Therefore, it is interesting to find out, if similar accumulation
of magnetization takes place in superconducting systems. It should
be noted that, despite formal similarities, such a magnetization
is fundamentally distinct from that induced by the normal SHE,
since it is not subject to the energy dissipation accompanying
spin diffusion and relaxation in normal systems.

We will consider SHE and the electric spin orientation for a
Josephson tunneling through a 2D normal contact (see Fig 1). The
SOI there is represented by the Hamiltonian
$H_{so}=\bm{\sigma}\cdot \mathbf{h}_{\bm{k}}$, where $\bm{\sigma}$
is a vector consisting of Pauli matrices. The spin-orbit field
$\mathbf{h}_{\bm{k}}$, which is a function of the electron wave
vector $\bm{k}$, can be given, for example, by Rashba
\cite{RashbaSOI}, or Dresselhaus \cite{Dresselhaus} SOI, as well
as by their combination. In this case the vector
$\mathbf{h}_{\bm{k}}$ lies in the plane of the 2D system. The
electron transport through the contact will be treated within the
diffusion approximation, so that the length of the junction $L$,
the electron coherence length $L_{c}$ and the spin precession
length $L_{so}=v_F/h$, where $v_F$ is the Fermi velocity and $h$
is the angular averaged spin orbit field, are assumed to be much
larger than the electron mean free path $l$. The electric voltage
across the junction is set to zero. Hence, the supercurrent is
provided by the phase difference between two electrodes. The
analysis of such a problem will be performed within a standard
semiclassical treatment of Gor'kov's equations in the diffusion
approximation (for a review see \cite{Efetov}). Our goal is to
derive linearized Usadel type equations and calculate the spin
density induced by SHE.

\begin{figure}[bp]
\includegraphics[width=3.8cm, height=4.5cm, angle=-90]{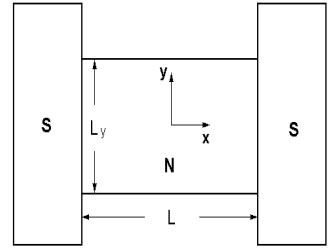}
\caption {Josephson contact with S and N denoting superconducting
and normal regions.} \label{fig1}
\end{figure}

As far as the thermal equilibrium state is considered, all
observables of interest can be expressed via retarded and advanced
Green functions. The corresponding Gor'kov's equations in the
Nambu representation have the form
\begin{equation}\label{G}
\left(i\frac{\partial}{\partial
t}-\check{H}-\check{\Sigma}^{r/a}\right)\check{G}^{r/a}(X,X^{\prime})=\delta(X-X^{\prime})\,,
\end{equation}
where $r$, $a$ denote retarded or advanced functions,
$X=t,\mathbf{r}$ and
\begin{equation}\label{H}
\check{H}=\frac{\tau_3}{2m^*}\hat{k}^2 -\tau_3
\mu+\bm{\sigma}\cdot \mathbf{h}_{\mathbf{\hat{k}}}\,,
\end{equation}
with the momentum operator $\mathbf{\hat{k}}=-i\partial/\partial
\bm{r}$ and the chemical potential $\mu$. After averaging of
initial Green functions over random positions of short-range
impurities, the self-energy in (\ref{G}) takes the form \cite{agd}
\begin{equation}\label{Sigma}
\check{\Sigma}^{r/a}(X,X^{\prime})=\frac{\tau_3}{2\tau\pi
N_F}\check{G}^{r/a}(t,t^{\prime},\bm{r},\bm{r})\tau_3\delta(\bm{r}-\bm{r}^{\prime})\,,
\end{equation}
where $\tau$ is the elastic scattering time. Unperturbed Green
functions are easily obtained from Eq.(\ref{G}). In the momentum
representation and after the time Fourier transform they can be
written as
\begin{equation}\label{G0}
\check{G}^{0r/a}(\omega,\bm{k})=\left(\omega-\tau_3
E_{k}-\bm{\sigma}\cdot \mathbf{h}_{\bm{k}}\pm
i\Gamma\right)^{-1}\,,
\end{equation}
where $E_k=(k^2/2m^*)-\mu$. Below we will perform calculations for
retarded functions and drop the labels $r,a$.

Proximity to superconducting contacts results in an admixture to
the Green functions of anomalous (proportional to $\tau_1$ and
$\tau_2$) components. Also, these functions  become inhomogeneous
in space. In order to calculate them, we will follow a well known
procedure in the framework of the semiclassical approximation
\cite{Landau}. First, we perform the Fourier transform with
respect to $X-X'$ introducing, accordingly, the frequency and wave
vector variables, $\omega$ and $\bm{k}$. The center of mass
variables will be remained intact and denoted as $\bm{r}$. Since
the problem is stationary, the corresponding center of mass time
variable is absent. Taking into account that variations of $G$ in
the scale of the Fermi wave-length are small, Eq. (\ref{G}) should
be expanded in terms of gradients $\partial/\partial \bm{r}$. The
next step is to simplify the self-energy part of Eq. (\ref{G})
keeping there only terms linear in the anomalous part. Such a
linearization can be done if the transparency of the SN contact is
small, or the leads are close to the superconducting critical
temperature. Taking the sum of Eq. (\ref{G}) and its conjugate
one, and making use of the fact that $\mathbf{h}_{\bm{k}}$ is an
odd function of $\bm{k}$, for the anomalous part $G_{12}$ we
obtain the equation
\begin{eqnarray}\label{G12}
(2\omega-\bm{v}\cdot\mathbf{\hat{q}}
+\frac{i}{\tau})G_{12}&-&\{\mathbf{h}_{\bm{k}}\cdot
\bm{\sigma},G_{12}\}- \nonumber \\
\frac{1}{2}\left[\delta\mathbf{h}_{\bm{k},\mathbf{\hat{q}}}\cdot
\bm{\sigma},G_{12}\right]&=&I_{sc}\,,
\end{eqnarray}
where
$\delta\mathbf{h}_{\bm{k},\mathbf{\hat{q}}}=(\mathbf{\hat{q}}\cdot\nabla_k)\mathbf{h}_{\bm{k}}$
with $\mathbf{\hat{q}}=-i\partial/\partial \mathbf{r}$, and
\begin{equation}\label{Isc}
I_{sc}= - \frac{1}{2\tau\pi N_F}\left( G^{0}_{11}g_{12} +
g_{12}G^{0}_{22}\right)\,.
\end{equation}
The lower labels in $G^0$ denote the matrix elements in the Nambu
space and $g_{12}=\sum_{\bm{k}}G_{12}$. Usually, the quantum
kinetic equation, such as (\ref{G12}), can be reduced to the
Eilenberger \cite{Eilenberger} equation by integration with
respect to the electron energy. In our case this procedure is not
convenient because of electron energy spin splitting. Instead,
within the diffusion approximation, from Eq. (\ref{G12}) we will
express $G_{12}$ in terms of $g_{12}$, and taking its sum over
$\bm{k}$ obtain the closed diffusion equation for $g_{12}$. Before
doing this, we transform the 2$\times$2 matrix
$G_{12,\alpha\beta}$ to the conventional pairing function
$F_{\alpha\overline{\beta}}\equiv G_{12,\alpha\beta}$, where
$\overline{\beta}$ denotes the spin projection opposite to
$\beta$. Further, it is convenient to decompose $F$ into triplet
$F_1, F_{-1}, F_0$ and singlet $F_s$ components as
\begin{eqnarray}\label{st}
F_{0}&=&\frac{F_{12}+F_{21}}{\sqrt{2}}\,,\,\,\,\,\,F_{s}=\frac{F_{12}-F_{21}}{\sqrt{2}}
\nonumber \\ F_1&=&F_{11}\,,\,\,\,\,\,\,\,\,\, F_{-1}=F_{22}.
\end{eqnarray}
The corresponding density function $f=\sum_{\bm{k}}F$ will also be
represented in a similar way. After this transformation, it is
easy to see that the last term in the l.h.s. of (\ref{G12}) is
responsible for a coupling between the singlet and triplet
components of the pairing function. Besides, the singlet-triplet
coupling also originates from the spin dependent parts of
$G^{0}_{11}$ and $G^{0}_{22}$ in Eq. (\ref{Isc}). Due to such
coupling, the triplet component of $F$ is generated within the
junction between two singlet superconductors.

For simplicity, when deriving the diffusion equation, let us
assume that SOI is strong enough, so that $L_{so} \ll L_c$.
Further, considering $I_{sc}$ together with the last term in the
l.h.s. of (\ref{G12}) as sources, we resolve Eq.(\ref{G12})
performing expansion in $(\bm{v}\cdot\mathbf{\hat{q}})\tau$ and
$\mathbf{h}_{\bm{k}}\tau$ up to the second order. Finally, we
obtain the following diffusion equation for the triplet pairing
function $f_m=(i/\pi N_F)\sum_{\bm{k}}F_m$, $(m=0,1,-1$):
\begin{equation}\label{diffusion}
2i\omega f=\tau\langle\left(-i
\bm{v}\cdot\frac{\partial}{\partial\bm{r}}+2\mathbf{J}\cdot
\mathbf{h}_{\bm{k}}\right)^2\rangle f +\mathrm{M} f_s\,,
\end{equation}
where $\mathbf{J}$ is the vector of 3$\times$3 angular moment
operators and $\langle...\rangle$ denotes the angular averaging
over the Fermi surface. The triplet-singlet coupling is given by
\begin{equation}\label{M}
\mathrm{M}_0=0\,\,,\mathrm{M}_{\pm1}=
\frac{4\tau^2}{\sqrt{2}}\langle
\mathrm{h}^{\mp}_{\bm{k}}(\mathbf{h}_{\bm{k}}\times\delta\mathbf{h}_{\bm{k},\mathbf{\hat{q}}})\rangle
f_s \,,
\end{equation}
with $\mathrm{h}^{\mp}_{\bm{k}}=\mathrm{h}^x_{\bm{k}}\mp
i\mathrm{h}^y_{\bm{k}}$. The singlet $f_s$ satisfies the usual
Usadel equation \cite{sfs} with an additional term which is
Hermitian conjugate to $M$. Since this term is small, we will
neglect a corresponding correction to $f_s$ in
Eq.(\ref{diffusion}). Hence, $f_s$ is given by the well known
unperturbed solution in the SNS contact. Since it varies within
the scale $L_c \gg L_{so}$, we neglected all contributions to
$\mathrm{M}$ with higher powers of gradients, as well as terms
proportional to $\omega \sim D/L^2_{c}$, where $D$ is the
diffusion constant.

Without the last term in the r.h.s., Eq.(\ref{diffusion}) formally
coincides with the spin diffusion equation for 2DEG in a zero
electric field \cite{MalshDiff}. The spin diffusion equation in
the presence of the electric field has been derived in Ref.
\cite{Misch} for the case of the Rashba SOI, and for a general SOI
in \cite{MalshAccumulation}. After a linear transformation
\cite{MalshDiff} to spin density variables Eq.(\ref{diffusion})
will also coincide with these equations, if, apart from a constant
factor, $f_s$ is formally identified with the electric field
potential. Hence, a coupling of the spin to the electric field in
normal spin transport appears to be very similar to the
singlet-triplet coupling in Eq.(\ref{diffusion}).

Let us consider an example of the Rashba SOI. In this case
$h^x_{\bm{k}}=\alpha k_y$ and $h^y_{\bm{k}}=-\alpha k_x$. For a
homogeneous in $y$-direction case all functions depend only on $x$
and we get $f_{0}=0,f_{1}=f_{-1}$, with $f_{1}$ satisfying the
equation
\begin{equation}\label{f1}
D\frac{\partial^2}{\partial x^2}f_1 - \Gamma_{so} f_1=
i\frac{\alpha\tau\Gamma_{so} }{\sqrt{2}}\frac{\partial}{\partial
x}f_s \,,
\end{equation}
where $\Gamma_{so}=2\tau\alpha^2 k_F^2$ is the D'yakonov-Perel'
spin relaxation time \cite{dp}. The small l.h.s. of
Eq.(\ref{diffusion}) has been neglected in (\ref{f1}). Boundary
conditions at $x=\pm L/2$ can be written in a way similar to a
singlet SN interface \cite{boundary}. At least in the linearized
approximation the boundary conditions contain only characteristics
of one-particle transmission. Therefore, they can be easily
generalized to the case of a triplet pairing. Following
calculations of Ref. \cite{boundary} we obtain
\begin{equation}\label{bc}
(f_{1S}- gf_{1N})\big |_{x=\mp L/2}=\pm b\frac{\partial
f_{1N}}{\partial x}\big |_{x=\mp L/2}\,,
\end{equation}
where the labels $S$ and $N$ denote superconductor and normal
sides of SN contacts at $x=\pm L/2$, and $g=|\omega
|/\sqrt{(\omega+i0^+)^2 -|\Delta|^2}$ is a DOS factor for a
superconductor. The characteristic length $b$ depends on the SN
barrier transmittance. For our choice of parameters $b \gg
L_{so}$. The same equation (\ref{bc}) takes place for $f_s$. At
the low SN barrier transmission one may use the so called rigid
boundary conditions and set $f_{1S}=0$. At the same time, the
singlet paring function $f_{sS}|_{\mp L/2}=g\Delta\exp(\pm
i\phi)/\omega$. Neglecting the third derivative of $f_s$, the
solution of Eq. (\ref{f1}) can be written as
\begin{equation}\label{psi}
f_1= -i\frac{\alpha\tau }{\sqrt{2}}\frac{\partial}{\partial x}f_s
+\psi(x)\,,
\end{equation}
where $\psi(x)$ is a linear combination of $\exp (\pm kx)$, with
$k=\sqrt{D/\Gamma} \equiv 1/L_{so}$. It is easy to see from
(\ref{bc}) that at $kb \gg 1$ the first term dominates in
Eq.(\ref{psi}). Therefore, $\psi$ will be neglected below.

Our next step is to calculate the spin polarization density
associated with triplet components of the pairing function. This
polarization is given by
\begin{eqnarray}\label{S}
S^i(\bm{r})&=&\frac{i}{2}\sum_{\bm{k}}\int
\frac{d\omega}{2\pi}n_F(\omega)\times \nonumber \\
&&Tr[\sigma^i(G^{r}_{\bm{k}11}(\omega,\bm{r})-G^{a}_{\bm{k}11}(\omega,\bm{r})]\,,
\end{eqnarray}
where $n_F$ is the equilibrium Fermi distribution function. It is
easy to see that the nonzero value of Eq. (\ref{S}) is provided by
triplet components of anomalous Green functions which contribute
to $G_{11}$ with a correction term $\propto f^2$. Up to the
leading second order with respect to $f_s$ and keeping only the
linear terms of the triplet $f_m$ $(m=1,-1, 0)$, for the retarded
function we obtain from Eqs. (\ref{G}-\ref{G0})
\begin{eqnarray}\label{trace}
\sum_{\bm{k}}Tr[\sigma^iG_{\bm{k}11}^{r/a}]&=&\frac{\mp 1}{\pi
N_F}[\frac{i\delta^{iz}}{2}(f_0^{r/a}
f^{+r/a}_s-f_s^{r/a}f_0^{+r/a}) +\nonumber
\\
&&\frac{1}{\sqrt{2}}(f_i^{r/a} f^{+r/a}_s+f_s^{r/a}f_i^{+r/a})]\,,
\end{eqnarray}
where $f_y=(f_1 + f_{-1})/2$ and $f_x=-i(f_1 -f_{-1})/2$. The
conjugate functions $f^+(\omega)=- f^*(-\omega)$.

In the case of Rashba SOI $f_x=f_0=0$ and $f_y=f_1$. The latter is
given by Eq. (\ref{psi}). Then, from (\ref{trace}) it immediately
follows that only the $y$-projection of the spin density is
finite. Using the relations $f^a_s(\omega)=f^r_s(-\omega)$ and
$f^a_m(\omega)=-f^r_m(-\omega)$ $(m=1,-1,0)$, we arrive to the
spin polarization
\begin{equation}\label{Sy}
S^y(x)=eN_F \alpha \tau \frac{J(x)}{\sigma_{dc}}\,,
\end{equation}
where $\sigma_{dc}$ is the dc conductivity of the normal metal and
$J$ is the Josephson current density
\begin{equation}\label{J}
J=\frac{e D}{4\pi^2 N_F}\int d\omega n_F(\omega)[(f_s^{r}
\frac{\partial f^{+r}_s}{\partial x} - \frac{\partial
f^{r}_s}{\partial x}f_s^{+r})-(r\rightleftarrows a)]\,.
\end{equation}
The  spin polarization (\ref{Sy}) coincides with polarization
induced in normal metals by the electric field $E$
\cite{Edelstein}, if the Josephson current is substituted for the
normal dissipative dc current $J_{dc}=\sigma_{dc} E$. It is easy
to check that this analogy takes place also for the Dresselhaus
SOI, with a little more complicated expression for $S^i(x)$
\cite{MalshAccumulation}. An important distinction from the
electric spin orientation in normal metals is that due to the
charge neutrality, $J_{dc}=const$ in the $x$ direction, while the
supercurrent varies inside the contact. Similar effect has been
predicted  by Edelstein \cite{Edelstein2} for bulk superconductors
and at NS boundary, providing the supercurrent flows along the SN
interface.

Let us now check, if the analogy with the electric spin
orientation extends to the spin-Hall effect. Hence, our goal is to
calculate $J^z_y$, which is the $y$ projection of a spin flux
polarized in the $z$-direction. The corresponding spin current
operator can be written as $\mathrm{J}^z_y=\{ \sigma_z,
\mathrm{v}_y \}/ 2$, where the velocity $\mathrm{v}_y=k_y/m^* +
\partial (\bm{\sigma}\cdot \bm{h}_{\bm{k}})/\partial k_y$. Since
it has been assumed that $h_z =0$, one gets
$\mathrm{J}^z_y=\sigma_z k_y/m^*$. The spin-Hall current $J_{sH}$,
in its turn, can be derived from Eq.(\ref{S}), with $\sigma^i$
substituted for $\mathrm{J}^z_y$. Keeping the same leading terms
as in calculation of the spin density, we arrive to $J_{sH}=0$.
This result does not depend on whether $\bm{h}_{\bm{k}}$ is given
by the Rashba or Dresselhaus interactions. That is very distinct
from the normal spin-Hall effect, where in the diffusive regime
the spin-Hall conductance is zero for the Rashba SOI, but finite
for the cubic Dresselhaus interaction \cite{MalshDress}. In
general, as it was discussed above, the zero value of $J_{sH}$ in
superconducting transport follows from the time inversion
symmetry.

Besides $J_{sH}$, in normal systems the DC current together with
SOI gives rise to accumulation of the $z$-component of spin at the
lateral edges of the sample
\cite{MalshAccumulation,accumulation,Bleibaum}. In the case of the
Josephson junction the $z$-projection of the spin density is given
by Eq.(\ref{S}) and the first term in Eq. (\ref{trace}). Hence, it
is proportional to the $f_0$ component of the pairing function
which, in its turn, can be found from Eq.(\ref{diffusion}). For
simplicity, let us consider hard wall boundaries of 2DEG at $y=\pm
L_y/2$. In this case one can borrow the boundary conditions for
Eq. (\ref{diffusion}) from Ref. \cite{MalshAccumulation,Bleibaum}.
In normal systems these conditions correspond to the vanishing
spin current at $y=\pm L_y/2$. In our case similar equations can
be written for triplet "currents" $\bm{j}=\sum_k \bm{v} F$. We
thus have $j^y|_{y=\pm L_y/2}=0$, where the 0 triplet component is
given by
\begin{equation}\label{j}
j^y_0=-D \frac{\partial f_0}{\partial y}  -2i\tau \langle(v_y
[\mathbf{h}_{\bm{k}}\times(\sqrt{2}\bm{f}+\tau
\delta\mathbf{h}_{\bm{k},\mathbf{\hat{q}}} f_s)] \rangle \,.
\end{equation}
The first term in this equation is the diffusive current, the
first term in the brackets is determined by the spin precession in
the effective spin-orbit field, and the last term looks as the
spin-Hall current in the normal spin transport. As far as $f_s$ is
treated a slowly varying function of $x$, thus allowing one to
ignore its higher gradients together with edge terms like $\psi$
in (\ref{psi}), the analysis of Eq. (\ref{diffusion}) with the
above boundary conditions is the same, as for SHE in normal
systems. Henceforth, following Ref.\cite{MalshAccumulation,
Bleibaum} one may conclude that $f_0=0$ for Rashba SOI, but $f_0$
is finite in the case of the cubic Dresselhaus interaction. From
Eqs.(\ref{S}) and (\ref{trace}) it is immediately seen that in the
former case $S^z=0$. For the Dresselhaus SOI the solution of Eq.
(\ref{diffusion}) has the form $f_0=\chi(y) (\partial f_s/\partial
x)$, where $\chi$ is a real odd function of $y$. Then,
Eqs.(\ref{S}),(\ref{trace}) and (\ref{J}) give  $S^z=-eN_F \chi(y)
(J/\sigma_{dc})$. The function $\chi$, in its turn, have been
calculated in Ref. \cite{MalshAccumulation}.

In conclusion, the spin-Hall effect induced by a supercurrent
across an SNS junction has been studied in the diffusive regime
for a relatively strong ($L_{so} \gg L_c$) SOI in the 2D junction
and for low conducting SN barriers. We found out that, although
the spin-Hall current is forbidden by the time inversion symmetry,
in the case of cubic Dresselhaus SOI the out-of-plane
magnetization is accumulated near sample edges at $y=\pm L_y/2$,
in a very close analogy to SHE in normal systems. Also, similar to
the electric spin orientation, the spin polarization parallel to
2DEG is finite throughout the entire N-region.

This work was supported by Russian RFBR, No 060216699, and Taiwan
NSC, No 96-2811-M-009-038.

\end{document}